\documentclass[graybox]{svmult}

\usepackage{mathptmx}       % selects Times Roman as basic font
\usepackage{helvet}         % selects Helvetica as sans-serif font
\usepackage{courier}        % selects Courier as typewriter font
\usepackage{type1cm}        % activate if the above 3 fonts are
\usepackage{makeidx}         % allows index generation
\usepackage{graphicx}        % standard LaTeX graphics tool
\usepackage{multicol}        % used for the two-column index
\usepackage[bottom]{footmisc}% places footnotes at page bottom

\makeindex             % used for the subject index
\begin{document}

\title*{Degrees of Freedom of the Quark Gluon Plasma, tested by Heavy Mesons}
\author{H. Berrehrah, M. Nahrgang, T. Song, V. Ozvenchuck,  P.B. Gossiaux, K. Werner, 
E.~Bratkovskaya  and J. Aichelin$^*$
%\footnote{invited speaker}
}
\institute{H. Berrehrah \at FIAS, University of Frankfurt, Ruth Moufang Str.1 60438  Frankfurt, Germany \and
M. Nahrgang
\at Department of Physics, Duke University, Durham, North Carolina 27708-0305, USA \and
T. Song \at FIAS, University of Frankfurt, Ruth Moufang Str.1 60438  Frankfurt \and
V. Ozenchuck \at IFJ PAN,Radzikowskiego 152, 31-342 Cracow, Poland \and
P.B. Gossiaux \at  SUBATECH, UMR 6457, Universit\'e de Nantes, Ecole des Mines de Nantes,
IN2P3/CNRS. 4 rue Alfred Kastler, 44307 Nantes cedex 3, France \and
K. Werner \at  SUBATECH, UMR 6457, Universit\'e de Nantes, Ecole des Mines de Nantes,
IN2P3/CNRS. 4 rue Alfred Kastler, 44307 Nantes cedex 3, France\and
E. Bratkovskaya \at GSI Helmholtzzentrum für Schwerionenforschung GmbH,
Planckstrasse 1,
64291 Darmstadt, Germany and Institut for Theoretical Physics, Johann Wolfgang Goethe Universit\"at, Max-von-Laue-Str. 1, 60438 Frankfurt am Main, Germany\and
$^*$J. Aichelin , invited speaker, \at SUBATECH, UMR 6457, Universit\'e de Nantes, Ecole des Mines de Nantes,
IN2P3/CNRS. 4 rue Alfred Kastler, 44307 Nantes cedex 3, France }
\maketitle

\abstract{Heavy quarks (charm and bottoms) are one of the few probes which are sensitive to the degrees of freedom of a Quark Gluon Plasma (QGP),
which cannot be revealed by lattice gauge calculations  in equilibrium. Due to the rapid expansion of the QGP energetic heavy quarks
do not come to an equilibrium with the QGP. Their energy loss during the propagation through the QGP medium depends strongly on 
the modelling of the interaction of the heavy quarks with the QGP quarks and gluons, i.e. on the assuption of the degrees of  freedom of the plasma.
Here we compare the results of different models,  the  pQCD based Monte-Carlo (MC@sHQ), the Dynamical Quasi Particle Model (DQPM) and the effective mass approach, 
for the drag force in a thermalized QGP  and discuss the  sensitivity of heavy quark energy loss on the 
properties of the QGP as well as on non-equilibrium dynamics.}

% as well as an application of those results 
%to the dynamical description of non-equilibrium situation occurring in heavy-ion
%collisions

\section{Introduction}
\label{sec:1}
The properties of infinite, strongly interacting systems in thermal equilibrium can presently only be determined by lattice gauge calculations.
In recent years the calculations of different groups converged    \cite{Borsanyi:2013bia, Bazavov:2014pvz} 
and therefore the pressure, the interaction measure and the entropy density as a function of temperature are known by now. These calculations predict that at high temperature and density the hadrons convert into a plasma of quarks and gluons (QGP). At  zero chemical potential the hadronic phase and the QGP phase are separated by a cross over. At finite chemical potentials, where presently the sign problem does not allow for lattice gauge calculations, the transition may be a first order phase transition as several QCD inspired models predict.  These lattice calculations do not reveal, however, the degrees of freedom in the QGP or at the transition between hadrons and QGP. The finite value of the interaction measure, $\epsilon - 3p$, were $\epsilon$ is the energy density and $p$ the pressure, which is zero for a noninteracting gas of Fermions and Bosons, tell us, however, that the constituents interact with each other but the kind of interaction remains unrevealed.  
On the other hand, these degrees of freedom are essential when we want to study the properties of the QGP beyond thermodynamics. They
influence the results if the QGP is tested by probes which do not come to a thermal equilibrium with the QGP. The other way round, the observables of these probes may reveal the degrees of freedom of the QGP or of hadrons close to the
phase transition. 

There is ample evidence that at top-RHIC and LHC energies  during ultrarelativistic heavy-ion collisions a color-deconfined QCD medium of high temperatures and densities, the quark-gluon plasma (QGP), is created. This allows for the first time
to study experimentally the properties of this  new state of matter, predicted by lattice gauge calculation. To study these properties one needs probes which do not come to a thermal equilibrium with the plasma particles, otherwise all their memory effects on the interaction with the plasma particles are lost.

One of the most promising probes are heavy-flavor quarks which are predominantly produced in the initial hard nucleon-nucleon interactions. Because these collisions are hard they can be calculated by perturbative QCD   \cite{FONLL,FONLL2,FONLL3}. Due to the propagation through the colored partonic medium high-$p_T$ heavy quarks suffer from a substantial energy loss, while low-$p_T$ heavy quarks are expected to thermalize at least partially within the medium.
The nuclear modification factor, $R_{\rm AA}$, which is the ratio of the spectra measured in heavy-ion collisions to the scaled proton-proton reference, and the elliptic flow, $v_2$, which is at low-$p_T$ a measure of thermalization inside the medium and reflects at high-$p_T$  the spatial anisotropy of the initial state, are presently the most discussed observables of heavy-flavor hadrons and their decay leptons.

 A suppression of high-$p_T$ D mesons, heavy-flavor decay electrons and muons has experimentally been measured by the STAR \cite{Stare,StarD} and Phenix \cite{Phenixe} collaborations at RHIC as well as the ALICE \cite{Alice,Dainese:2012ae,Abelev:2013lca} and CMS \cite{CMS:2015hca} collaborations at LHC. Finite values of $v_2$ of $D$ mesons, heavy-flavor decay electrons and muons was found  both at RHIC \cite{Adamczyk:2014yew} and at LHC \cite{Adam:2015pga}.

Perturbative QCD calculations for the average energy loss of high-$p_T$ particles include elastic \cite{elastic,Peshier:2006hi,Gossiaux:2008jv,Peigne:2008nd} and/or inelastic scatterings \cite{radiative,Wang95,Baier95,Baier97,Zakharov,GLV,Dokshitzer,AMY,ASW,Zhang:2003wk,Djordjevic:2003zk,Djordjevic:2004nq,Djordjevic:2007at,Djordjevic:2009cr}. In most of these models, no evolution of the QGP is considered and only average temperatures and path-length distributions are included. The generic form of the $R_{\rm AA}$ as a function of $p_T$ or the integrated $R_{\rm AA}$ as a function of centrality can easily be reproduced by most calculations on the basis of fundamental principles despite rather different ingredients. The strength of the suppression, however,  depends  strongly on the details of the space-time evolution of the QGP \cite{Renk:2011aa}. For quantitative predictions  the fully coupled dynamics of the heavy quarks and of the QGP needs to be taken into account. Therefore we concentrate here on three approaches which have in common that they use
not only a Boltzmann collision kernel to describe the interaction of the heavy quark with the plasma particle but as well a dynamical time evolution of the QGP itself. 

1) The first of these models is the  MC@sHQ approach which assumes that gluons and quarks of the QPG are massless. The interaction of the heavy quark with the plasma particles uses Born type diagrams with a coupling constant depending on the momentum transfer and a hard thermal loop inspired gluon propagator. Here two versions are available, one in which the heavy quarks interact only elastically with the QGP particles and one which includes in addition radiative collisions
(i.e. gluon bremsstrahlung).
The expansion of the QGP is described by the EPOS event generator. 

2) The second approach, dubbed effective mass approach assumes that the gluons and quarks in the entrance and exit channel of the elementary interactions are massive. Their mass is obtained by a fit to the entropy density calculated by lattice gauge calculations. 

3) The third model is the  PHSD approach which uses the dynamical quasi particle mode (DQPM) to calculate the masses and widths of the plasma constituents as well as temperature dependent coupling constants from a fit to the results of lattice data. Collisions between heavy quarks and light quarks and gluons are here limited to elastic collisions. They are calculated by Born diagrams with "re-summed" propagators and vertexes.

In section 2 we start out with the description of the  MC@sHQ approach, section 3 is devoted to the models which treat quarks and gluons as quasiparticles. In section 4 we introduce the drag coefficient and discuss the results obtained for the different models.

\section{The standard MC@sHQ approach}
In the standard  MC@sHQ approach \cite{Gossiaux:2008jv,Aichelin:2013mra} the heavy quarks can interact with the
plasma constituents purely elastically or in a combination of elastic and inelastic  collisions. The elastic cross sections in Born approximation are obtained within a hard thermal loop (HTL) calculation, including a running coupling constant $\alpha_s$ \cite{Peshier:2006ah,Gossiaux:2008jv}. The contribution from the $t$-channel is regularized by a reduced Debye screening mass $\kappa m_D^2$, which is calculated self-consistently \cite{Gossiaux:2008jv,Peigne:2008nd}, yielding a gluon propagator with
\begin{equation}
 1/Q^2\to1/(Q^2 -\kappa \tilde{m}_D^2(T))
\label{prop}
\end{equation}
 for a  momentum transfer $Q^2$. In this HTL+semihard approach \cite{Gossiaux:2008jv}, $\kappa$ is determined such that the average energy loss is maximally insensitive to the intermediate scale between soft (with a HTL gluon propagator) and hard (with a free gluon propagator) processes. The inelastic cross sections include both, the incoherent gluon radiation \cite{Gunion:1981qs} and the effect of coherence, i.e. the Landaul-Pomeranchuk-Migdal (LPM) effect \cite{Gossiaux:2012cv}. In this approach the incoming light partons are considered as massless. Results for heavy-flavour observables previously obtained within MC@sHQ+EPOS2 \cite{Nahrgang:2014vza,Nahrgang:2014ila,Nahrgang:2013xaa,Nahrgang:2013saa} considered massless light partons in the QGP.

The fluid dynamical evolution is used as a background providing us with the temperature and velocity fields necessary to sample thermal scattering partners for the heavy quarks. 
The  MC@sHQ approach couples the Monte-Carlo treatment of the  Boltzmann equation of heavy quarks (MC@sHQ) \cite{Gossiaux:2008jv} to the $3+1$ dimensional fluid dynamical evolution of the locally thermalized QGP following the initial conditions from EPOS2 \cite{Werner:2010aa,Werner:2012xh}. EPOS2 is a multiple scattering approach which combines pQCD calculations for the hard scatterings with Gribov-Regge theory for the phenomenological, soft initial interactions. Jet components are identified and subtracted while the soft contributions are mapped to initial fluid dynamical fields. By enhancing the initial flux tube radii viscosity effects are mimicked, while the subsequent $3+1$ dimensional fluid dynamical expansion itself is ideal. Including final hadronic interactions the EPOS2 event generator has successfully described a variety of bulk and jet observables, both at RHIC and at LHC \cite{Werner:2010aa,Werner:2012xh}.  For details we refer to the references.

\begin{figure}[tb]
\centering
\includegraphics[width=0.48\textwidth]{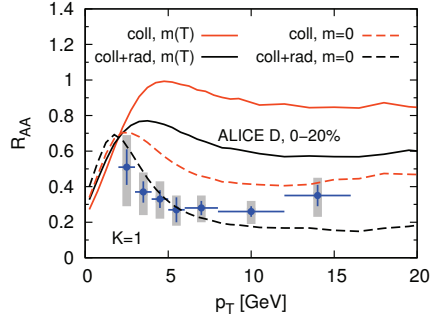}
\caption{(Color online) Comparison of the $D$ meson $R_{\rm AA}$ for a QGP consisting of massive quasiparticles (solid lines) and massless partons (dashed lines). Purely collisional (orange, light) and collisional+radiative(LPM) (black
line) energy loss scenarios are shown.}
 \label{fig:K1}
\end{figure}

Including elastic and inelastic collisions this approach reproduces quite well the experimental D-meson and non photonic electron data at RHIC and LHC. As an example we display in Fig. \ref{fig:K1} the $D$ meson $R_{\rm AA}$ as dashed line, for elastic (coll) as well as for elastic+inelastic collisions (coll+rad) in comparison with ALICE data. Elastic cross sections alone give not sufficient stopping in this approach.

\section{Quarks and Gluons as Quasiparticles}
\label{sec:2}
It is well known that quasiparticle models are able to reproduce the lattice QCD equation of state \cite{ Cassing:2009vt,Bluhm:2004xn,Berrehrah:2013mua} by assuming effective dispersion relations for noninteracting quasi-quarks and -gluons in the QGP. Due to the statistical factor of $\exp[-m/T]$ we expect that in a medium with a given temperature the density of light massive partons is reduced as compared to the density of massless partons, what leads to a reduced scattering rate. Thus the mass of the plasma constituents has an immediate influence on the stopping of energetic heavy quarks during their passage through the QGP, or the other way around, measuring the stopping of heavy quarks allows to conclude on the properties of the quasi particles in the medium. 
\subsection{The effective mass approach}
\label{sec:3}
Our second approach is an  extension \cite{Nahrgang:2016lst} of  the model established in \cite{Gossiaux:2008jv} by assuming  that the incoming and outgoing  light partons, which interact with the heavy quarks, have a finite mass. For this purpose, we treat those as well as  long-living quasiparticles. 
\begin{figure}[h]
 \centering
 \includegraphics[width=0.45\textwidth]{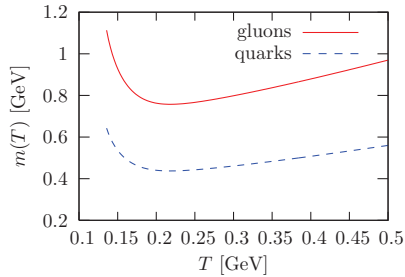}
 \caption{(Color online) Thermal masses of the quarks and gluons in the QGP within an effective mass approach.}
 \label{fig:thmass}
\end{figure}

The temperature dependence of the parton masses is obtained from fitting the entropy density of a noninteracting gas of massive quarks and gluons to the lattice equation of state \cite{Borsanyi:2013bia,Bazavov:2014pvz}.

The pressure and the energy density read
\begin{eqnarray}
p(T)&=&d_q\int\frac{{\rm d}^3p}{(2\pi)^3}\frac{p^2}{3 E_q}f_{\rm FD}(E_q) \nonumber \\
     &&+ d_g\int\frac{{\rm d}^3p}{(2\pi)^3}\frac{p^2}{3 E_g}f_{\rm BE}(E_g)-B(T)
\label{eq:pthmass}
\end{eqnarray}
\begin{eqnarray}
e(T)&=&d_q\int\frac{{\rm d}^3p}{(2\pi)^3} E_q f_{\rm FD}(E_q)\nonumber  \\
     &&+ d_g\int\frac{{\rm d}^3p}{(2\pi)^3} E_g f_{\rm BE}(E_g)+B(T)
\label{eq:ethmass}
\end{eqnarray}
with $E_q=\sqrt{p^2+m_q^2}$, $E_g=\sqrt{p^2+m_g^2}$ and the temperature dependent bag constant $B(T)$. $f_{\rm FD}$ and $f_{\rm BE}$ are the Fermi-Dirac and Bose-Einstein distributions respectively. In order to connect $m_q$ and $m_g$ we use the perturbative HTL-result $m_g=\sqrt{3} m_q$ \cite{Thoma:1995ju} as a conservative estimate. We assume the same thermal masses for $u$, $d$ and $s$ quarks.
The mean-field contribution $B$ cancels in the entropy density
\begin{equation}
s(T)=\frac{e(T)+p(T)}{T}\, .
\label{eq:entropyfit}
\end{equation}

 The thermal masses of quarks and gluons, obtained by this procedure, are shown in Fig. \ref{fig:thmass}. At high temperatures we find the almost linear behavior as it is known from pQCD calculations. The quasiparticle masses show a strong increase for temperatures above and close to $T=134$~MeV, which coincides very well with the effective transition temperature $T_f$ from the EPOS parametrization, the transport model which describes the expansion of the QGP in this approach. In this simple quasiparticle picture no assumption about the functional form of the temperature dependence of the thermal masses is made. Other quasiparticle approaches \cite{Peshier:2002ww,Bluhm:2006yh} but also the DQPM model, discussed in the next subsection, express the masses via the perturbative form $m^2\propto g^2T^2$ and parametrize a logarithmic temperature-dependence of the coupling $g$ by a fit to the lattice QCD equation of state.  The definition of the running coupling constant at finite temperatures is not unique. In the effective mass approach one does not assume any explicit temperature dependence of  $\alpha_s$. 
The coupling is determined by the momentum transfer in the individual scattering process.

A finite mass of the light partons reduces substantially the particle density at a given temperature. Therefore the heavy quarks have less scattering partners and the scattering rate is reduced. A lower scattering rate translates directly into a lower energy loss
as can be seen in fig. \ref{fig:K1} where the full lines represent the results for the effective mass approach for the same time evolution of the plasma as for the standard  MC@sHQ approach.

\subsection{The Dynamical QuasiParticle Model (DQPM)}

The DQPM describes QCD properties in terms of 'resummed' single-particle Green's functions (in the sense of a two-particle irreducible (2PI) approach). In other words: the degrees-of-freedom of the QGP are  interpreted as being strongly interacting massive effective quasi-particles with broad spectral functions (due to the high interaction rates). The dynamical quasiparticle entropy density $s^{DQP}$ has been fitted to lattice QCD calculations which allows to fix for $\mu_q=0$ the 3 parameters of the DQPM entirely (we refer to the Refs. \cite{Cassing:2009vt,Cassing:2008nn,Cassing:2007yg} for the details of the DQPM model).

The DQPM employs a Lorentzian parametrization of the partonic spectral functions $A_{i} (\omega_{i})$,
where $i$ is the parton species:
{%\setlength\arraycolsep{-10pt}
\begin{eqnarray}
\label{equ:Sec2.2}
A_i (\omega_i) & & \ = \frac{\gamma_i}{\tilde{E}_i} \biggl(\frac{1}{(\omega_i - \tilde{E}_i)^2 + \gamma_i^2} -
\frac{1}{(\omega_i + \tilde{E}_i)^2 + \gamma_i^2} \biggr)
\nonumber\\
& & {} \equiv \frac{4 \omega_i \gamma_i}{(\omega_i^2 - {\mathbf p}_i^2 - M_i^2)^2 +
4 \gamma_i^2 \omega_i^2},
\end{eqnarray}}
with $\tilde{E}_i^2 ({\mathbf p}_i) = {\mathbf p}_i^2 + M_i^2 - \gamma_i^2$, and $i \in [g, q, \bar{q}, Q, \bar{Q}]$.
The spectral functions $A_i (\omega_i)$ are normalized as:
{\setlength\arraycolsep{0pt}
\begin{eqnarray}
%%%\label{equ:Sec2.3}
\int_{- \infty}^{+ \infty} \frac{d \omega_i}{2 \pi} \ \omega_i
\ A_i (\omega_i, {\mathbf p}) = \!\!\! \int_0^{+ \infty} \frac{d \omega_i}{2 \pi} \ 2 \omega_i \ A_i (\omega_i, {\mathbf p}_i) =
1 , \nonumber
\end{eqnarray}}
where $M_i$, $\gamma_i$ are the dynamical quasi-particle mass (i.e. pole mass) and width of the
spectral function for particle $i$, respectively. They are directly related to the real and imaginary parts of the related self-energy, e.g. $\Pi_i = M_i^2 - 2 i \gamma_i \omega_i$, \cite{Cassing:2009vt}. In the off-shell approach, $\omega_i$ is an independent variable and related to the \emph{``running mass''} $m_i$ by: $\omega_i^2 = m_i^2 + {\mathbf p}_i^2$. The mass (for gluons and quarks) is assumed to be given by the thermal mass in the asymptotic high-momentum regime. We note that this approach is consistent with respect to microcausality in field theory \cite{Rauber}.
\begin{figure}[h]
 \centering
 \includegraphics[width=0.5\textwidth]{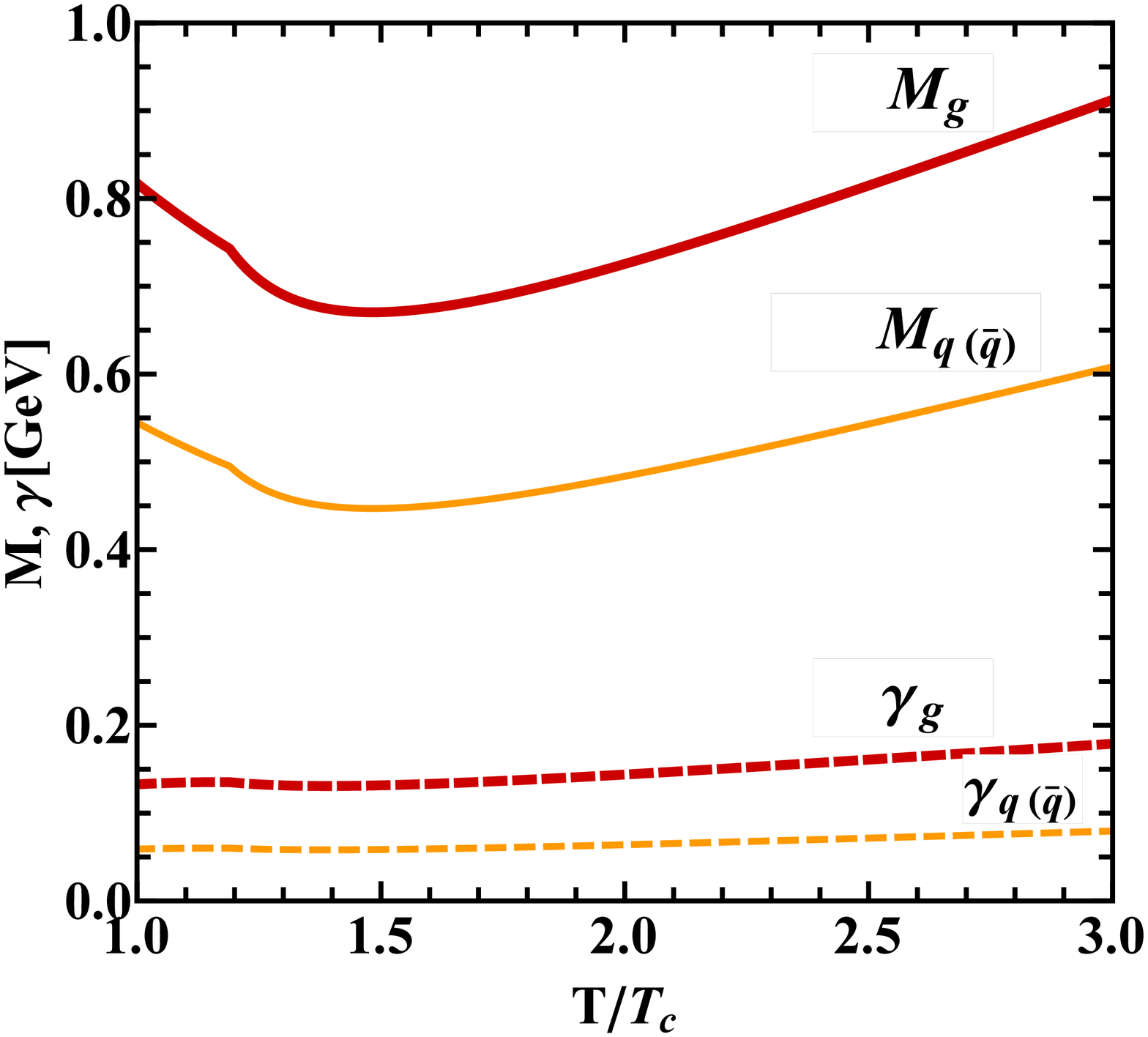}
 \includegraphics[width=0.45\textwidth]{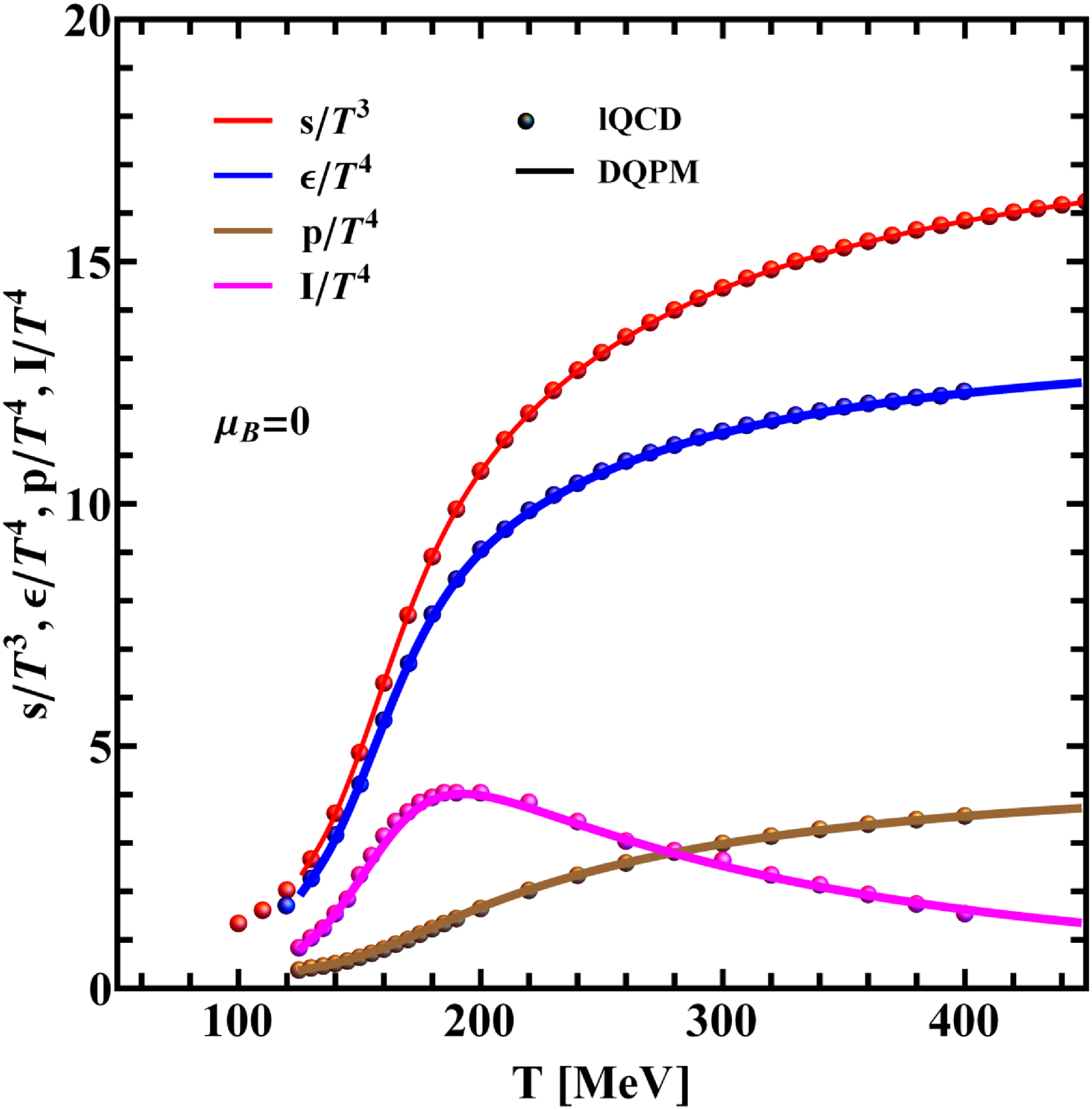}
\caption{(Color online) Left: The effective gluon mass
$M_g$ and witdh $\gamma_g$ as function of the scaled temperature $T/T_c$ (red lines).
The blue lines show the corresponding quantities for quarks. Right: The scaled entropy density $s(T)/T^3$ (blue line) and scaled energy
density $\epsilon(T)/T^4$ (red line) from the DQPM in comparison to
the lQCD results
% from Ref. \cite{aori10} 
(full dots and triangles). }
 \label{fig:thmassdqpm}
\end{figure}
The DQPM model has originally been designed to reproduce the QCD equation of state, calculated on the lattice, at zero chemical potential $\mu_q$ in an effective quasiparticle approach. The fit to the lattice data yields the masses
\begin{eqnarray}
\label{equ:Sec2.6}
& & M_g^2 (T) = \frac{g^2 (T/T_c)}{6} \Biggl((N_c + \frac{1}{2} N_f) T^2  \Biggr) \ ,
\nonumber\\
& & {} M_q^2 (T) = \frac{N_c^2 - 1}{8 N_c} g^2 (T/T_c) \Biggl( T^2 
\Biggr)\ ,
\end{eqnarray}
and the widths
\begin{eqnarray}
\label{equ:Sec2.9}
&&\hspace*{-0.78cm}\gamma_g (T)\!=\!\frac{1}{3} N_c \frac{g^2 (T/T_c)}{8 \pi} \, T \ln \!\left(\!\frac{2 c}{g^2(T/T_c)}+1\! \right)
\nonumber\\
&&\hspace*{-0.78cm}\gamma_q (T)\!=\!\frac{1}{3}\frac{N_c^2 - 1}{2 N_c}\frac{g^2 (T/T_c)}{8 \pi}\,T \ln \!\left(\!\frac{2 c}{g^2 (T/T_c)}+1 \!\right).\end{eqnarray}
The masses and widths as a function of the scaled temperature are displayed in Fig. \ref{fig:thmassdqpm} (left).  We see that the mass of the quasiparticles has a minimum around $1.5\ T_c$ and increases at lower and higher temperatures where the increase is linear corresponding to the perturbative thermal Debye mass. 
The last fit parameter is the coupling constant for which on obtains
\begin{eqnarray}
\label{equ:Sec2.6}
& & \hspace{-0.1cm} \displaystyle{g^2 (T/T_c) = \frac{48 \pi^2}{(11 N_c - 2 N_f) \ln \left(\lambda^2 (\frac{T}{T_c} - \frac{T_s}{T_c})^2 \right)}  \hspace{0.3cm}  T > T^{\star} = 1.19 \ T_c},
\nonumber\\
& & {} \hspace{-0.1cm} \displaystyle{g^2 (T/T_c) \rightarrow g^2 (T^{\star}/T_c)
\left(\frac{T^{\star}}{T} \right)^{3.1}  \hspace{0.7cm}  T < T^{\star} = 1.19 \ T_c.}
\end{eqnarray}
%\begin{equation}
%\hspace*{-0.78cm}\displaystyle{g^2( T/T_c)\!=\!\!\frac{48 \pi^2}{(11 N_c - 2 N_f) \ln\!\left(\!\lambda^2 (\frac{T}{T_c} - %\frac{T_s}{T_c})^2\!\right)}}
%label{coup}
%\end{equation}
with $\lambda = 2.42$, c=14.4  and $T_s = 73$  MeV. Eqs. (\ref{equ:Sec2.6}) and (\ref{equ:Sec2.9}) define the DQPM ingredients necessary for the calculations at finite temperature. In this article we limit ourselves to on-shell quarks and gluons because we have found in Ref.  \cite{Berrehrah:2014kba} that a finite width $\gamma_{g,q,Q}$ for gluons $g$, light $q$ and heavy quarks $Q$ has an impact of about 10-20\% on the heavy-quark  transport coefficients. With this set of fit parameters one obtains an excellent reproduction of the lattice data as one can see in Fig. \ref{fig:thmassdqpm}, right.
\begin{figure}[h]
\centerline{
\includegraphics[width=0.65\textwidth]{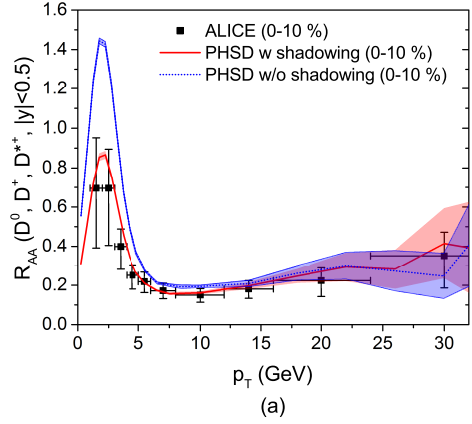}}
\caption{(Color online) The ratio  $R_{\rm AA}$ of $D^0,~D^+$, and $D^{*+}$ mesons
within $|y|<0.5$ as a function of ${\rm p_T}$ in 0-10 \% 
central Pb+Pb collisions at $\sqrt{s_{\rm NN}}=$2.76 TeV  \cite{Song:2015ykw} compared with the
experimental data from the ALICE collaboration
%~\cite{Adam:2015sza}. 
The solid and
dotted lines are, respectively, $R_{\rm AA}$ with and without (anti-)shadowing.
The charm quark mass is taken to be 1.5 GeV.}
\label{raa0}
\end{figure}

Based on the Kadanoff Baym equations, with these ingredients a transport theory, the parton hadron string dynamics (PHSD), has been developed which can describe a multitude of observables in ultrarelativistic heavy ion collisions. In particular these masses and coupling constants enter the Boltzmann collision integral in which the scattering is treated in Born approximation.  For details we refer
again to the Refs \cite{Cassing:2009vt,Cassing:2008nn,Cassing:2007yg}. This model describes the heavy quark observables at RHIC \cite{Song:2015sfa} and LHC \cite{Song:2015ykw} energies. As an example we display in Fig. \ref{raa0} the calculation for Pb+Pb in comparison with the experimental data {\bf at the LHC}.

\section{The Drag coefficient}
How can one compare three models, which have a multitude of different ingredients and which give nevertheless quite similar results when compared to the experimental data?  The comparison of cross sections themselves (which are a function of the momentum transfer, of the momentum of the scattering partner, of the temperature of the QGP and of the different channels which are considered) is not sufficient since one needs to know which temperatures and momentum transfers are important for the time evolution of the QGP.  As a first step it is useful to assume an equilibrium situation and to compare transport coefficients. To understand the meaning of the drag force, the transport coefficient which we study here, it is best to start out from the assumption that the time evolution of the heavy-quark distribution
function, $f(\vec{p},t)$, in the QGP can be described by a
Fokker-Planck/Langevin approach \cite{Svetitsky:1987gq,Moore:2004tg,vanHees:2004gq,vanHees:2005wb,Greco:2007sz,He:2011qa,Cao:2013ita},  
\begin{equation}
\frac{\partial f(\vec{p},t)}{\partial t}
=\frac{\partial}{\partial p_i}\left[ A_i(\vec{p})f(\vec{p},t)+
\frac{\partial}{\partial p_j}B_{ij}(\vec{p})f(\vec{p},t)\right]\, .
\end{equation}
The interaction of a heavy quark with the QGP is expressed by a drag force $A_i$ and a diffusion tensor $B_{ij}$, which can be written as $B_\perp$ and $B_{||}$. These quantities can be calculated from the microscopic $2\rightarrow 2$ processes by 
  \begin{eqnarray}
   \frac{{\rm d}X}{{\rm d}t} &=& \frac{1}{2E}
      \int \frac{d^3k}{(2\pi)^3 2 E_k}
      \int \frac{d^3k'}{(2\pi)^3 2E_{k'}}\,
      \int \frac{d^3p'}{(2\pi)^3 2E'}\, \nonumber \\
 &&      \times\, \sum \frac{1}{d_i}
  \left|{\cal M}_{i,2\to2}\right|^2\, n_i(k)\,  X \nonumber \\
      &&      \times\, (2\pi)^4\delta^{(4)}(p\!+\!k\!-\!p'\!-\!k')\, , 
  \label{eq:x}
  \end{eqnarray}
where $p(p')$ and $E=p_0$ ($E'=p_0')$ are momentum and energy of the heavy quark before (after) the collision and $k(k')$ and $E_k=k_0$ ($E_{k'}=k_0')$ are momenta and energies of the colliding light quark ($i=q$) or gluon ($i=g$). For the scattering process of a heavy quark with  a light quark ($qQ\to qQ$) $d_q=4$ and for the scattering off a gluon ($gQ\to gQ$) $d_g=2$. $n(k)$ is the thermal distribution of the light quarks or gluons. ${\cal M}_i$ is the matrix element for the scattering process $i$, calculated using pQCD Born matrix elements. 
In order to calculate the quantities mentioned above, $A_i$ and $B_{ij}$, one has to take $X=p-p'_i$ and $X=1/2(p-p'_i)(p-p'_j)$.
\begin{figure}
 \centering
 \includegraphics[width=0.48\textwidth]{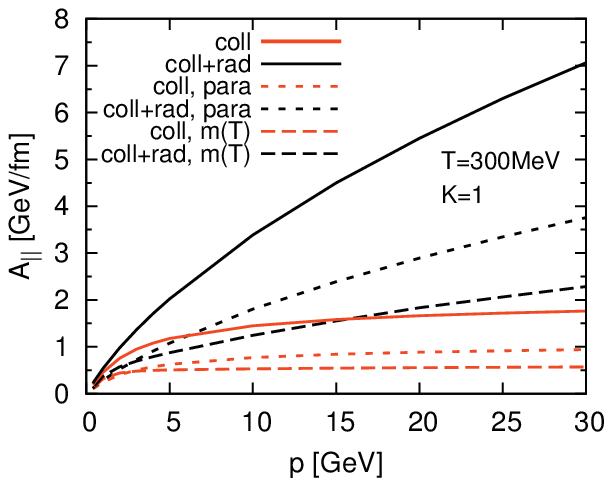}
\includegraphics[width=0.48\textwidth]{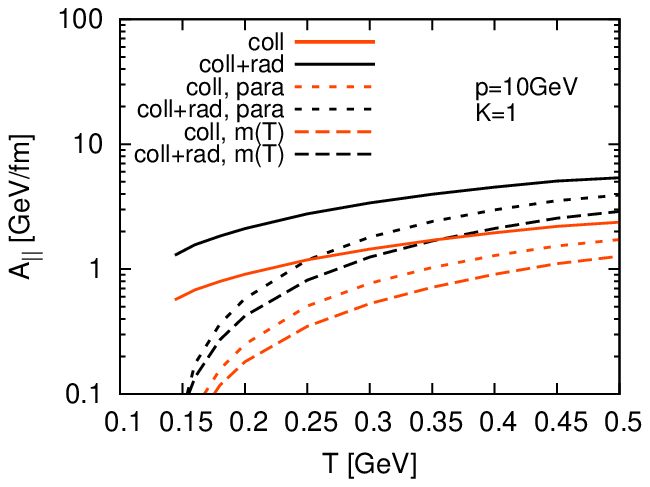}
 \caption{(Color online) The drag force  of c-quarks in the plasma rest frame  for three different representations of the QGP constituents: massless partons (solid), EPOS parametrization of the equation of state (short dashed) and massive quasiparticles (long dashed), as a function of the momentum (left) and as a function of the medium temperature (right) \cite{Nahrgang:2016lst}. Results for the purely collisional energy loss (black) and for the collisional+radiative(LPM) are shown.}
 \label{fig:AdragnoK}
\end{figure}

\begin{figure}[h!] %tbh!
\begin{center}
\includegraphics[width=0.45\textwidth]{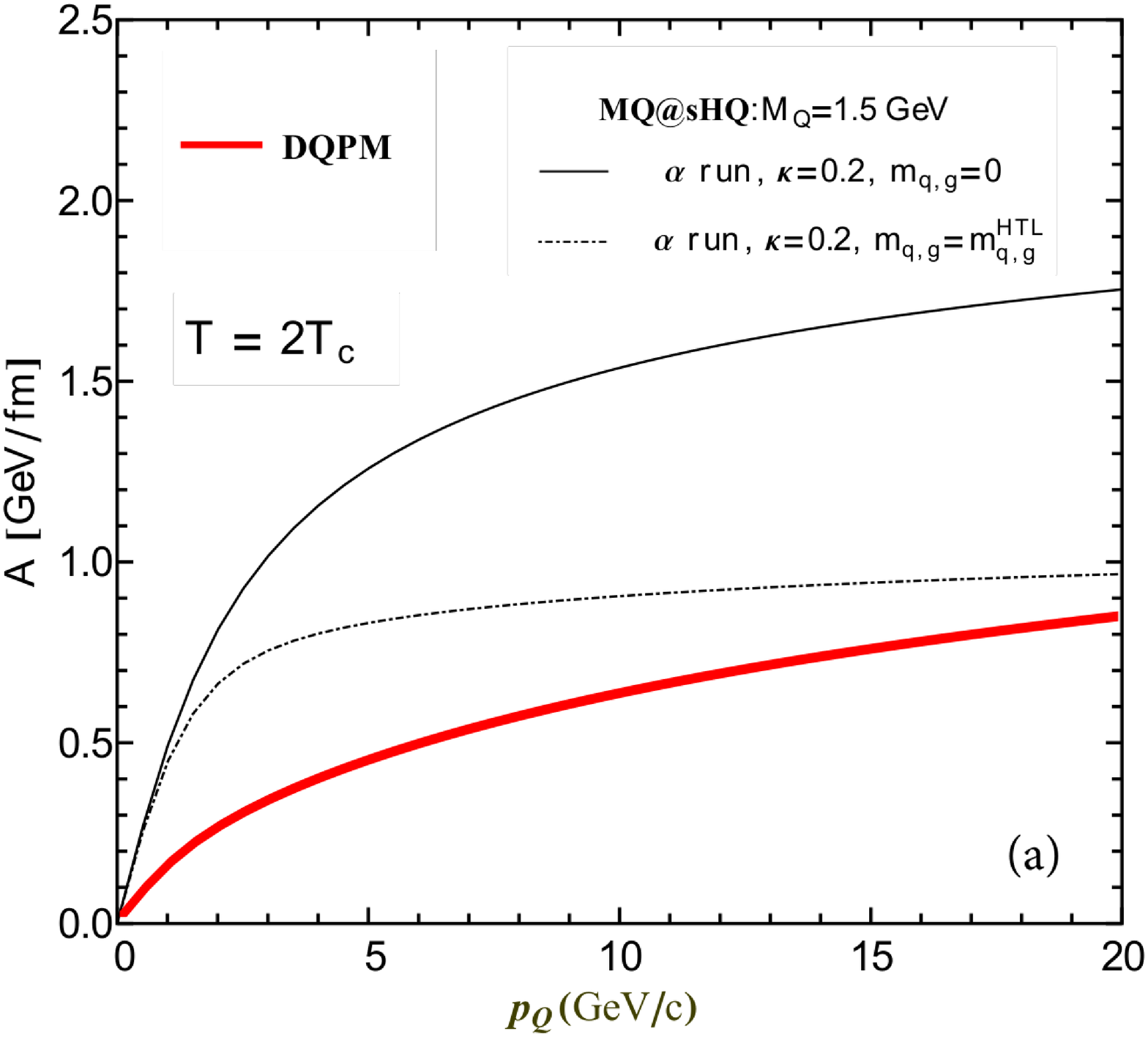}
\includegraphics[width=0.45\textwidth]{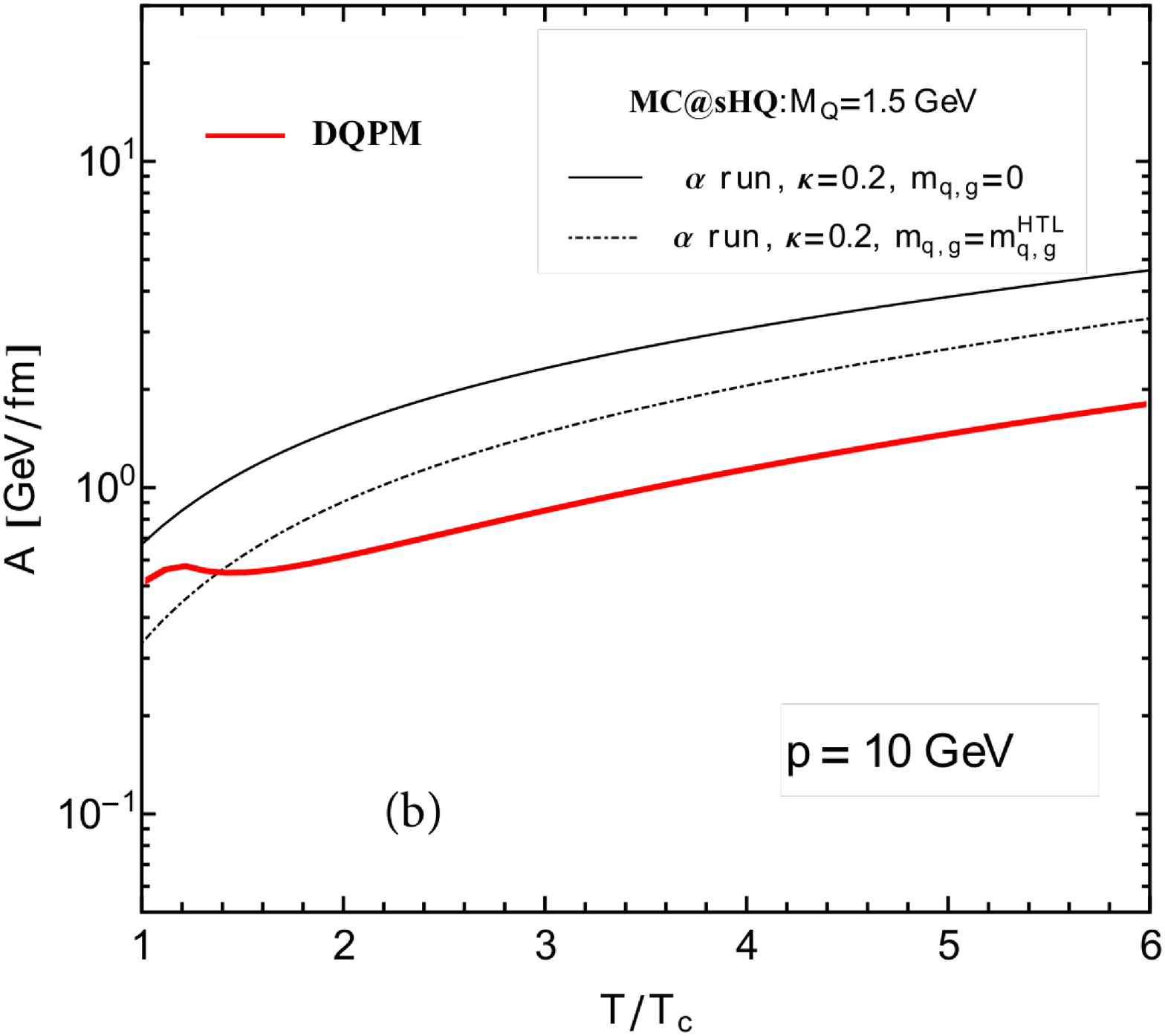}
%\vspace{-0.2cm}
\caption{{(Color online) The drag force $A$ of c-quarks in the plasma rest frame 
for three different approaches as a function of the heavy quark momentum $p_Q$ for $T = 2 T_c$  (a) and  as a function of the
temperature $T/T_c$ ($T_c = 0.158$ GeV) for an intermediate heavy quark momentum, $p_Q = 10$ GeV (b) \cite{Berrehrah:2014kba}.}}
\label{fig:cDragDifferentModel}
\end{center}
\end{figure}

Usually, the simultaneous calculation of both coefficients from Eq. \ref{eq:x} does not satisfy the Einstein relation which assures that asymptotically $ f(\vec{p},t)$ is the distribution function at thermal equilibrium. In most Fokker-Planck/Langevin approaches one quantity is calculated and the other one is obtained via the Einstein relation under the assumption that $B_\perp= B_{||}$. It has recently been shown that the results from the Fokker-Planck/Langevin approach differ substantially from that of the Boltzmann equation in which the collision
integrals are explicitly solved \cite{Das:2013aga} because the underlying assumption, that the scattering angles and the momentum transfers are small, is not well justified. A recent review article \cite{Andronic:2015wma} gives a broad overview over the various approaches of heavy-flavor energy loss using either the Fokker-Planck/Langevin or the Boltzmann dynamics.
\begin{figure}[h]
%\sidecaption[t]
\centering
\includegraphics[width=0.48\textwidth]{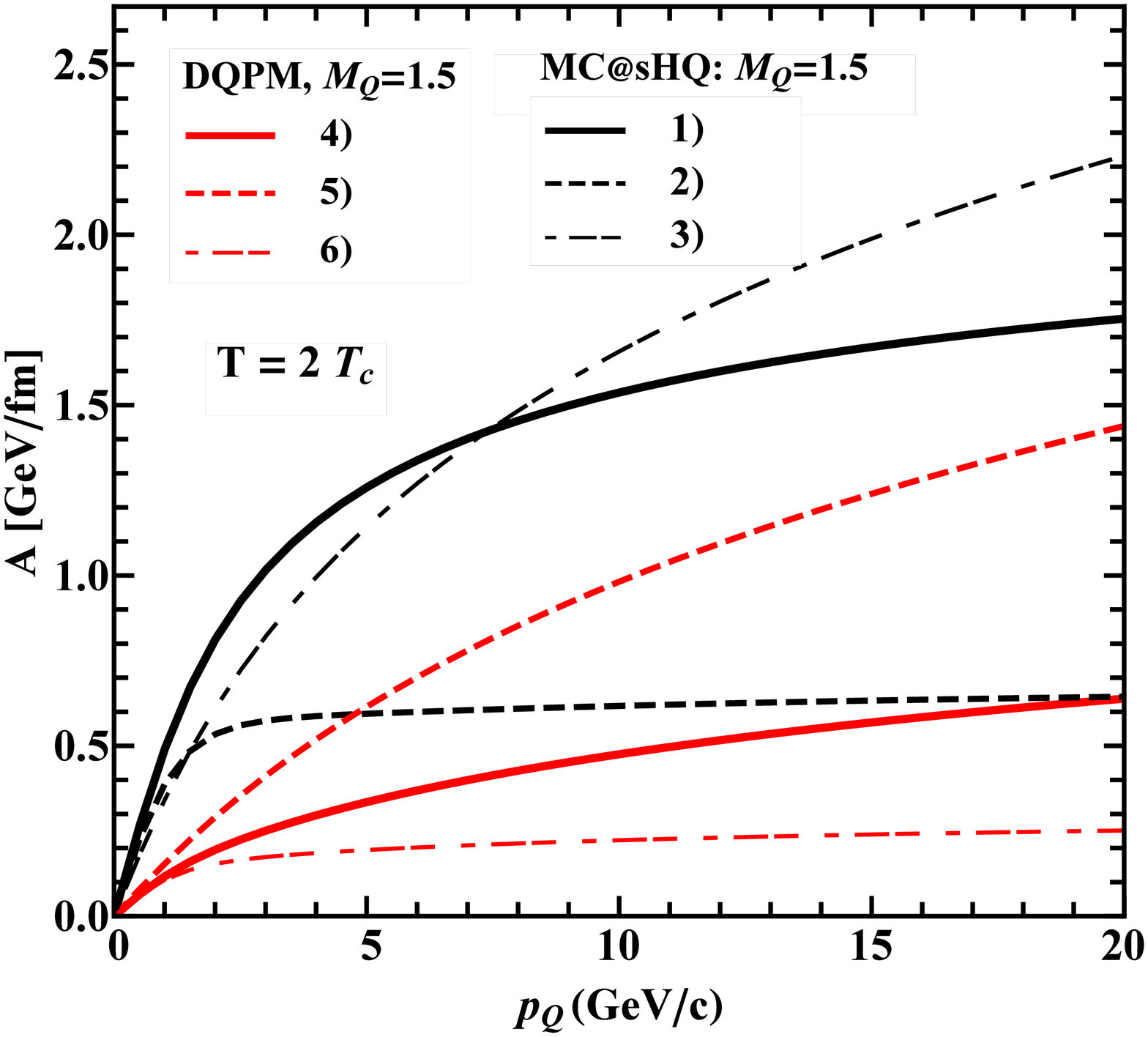}
\includegraphics[width=0.48\textwidth]{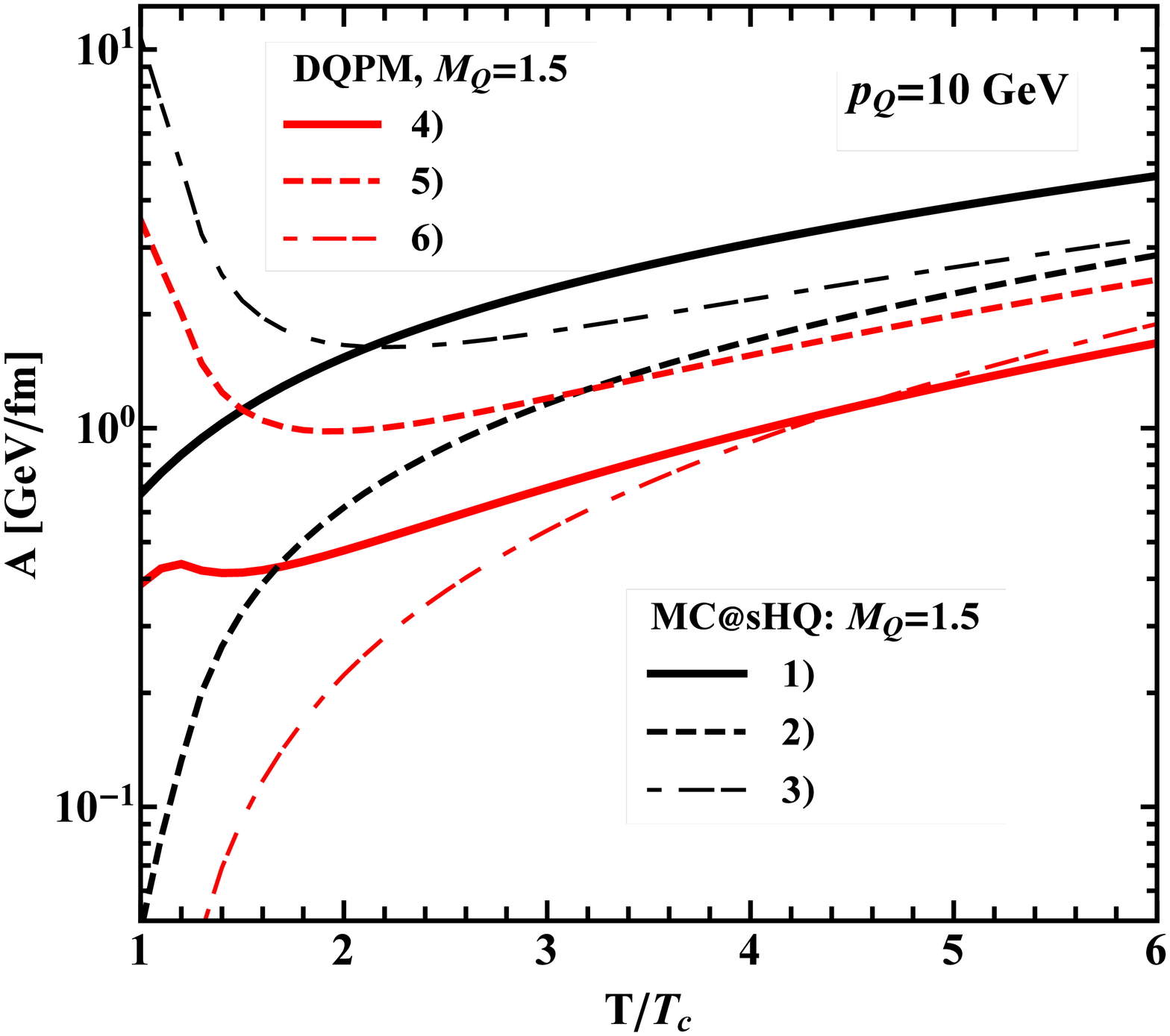}
\caption{Drift coefficient A of a heavy quark in a QGP of $ T= 2 T_c$  for different assumptions of particle masses and coupling constants (see text and Table 1).}
\label{fig:5}       % Give a unique label
\end{figure}
\begin{table}[h]
\centering
\begin{tabular}{|c|c|c|l|}
\hline
&coupling &	mass in gluon propagator & \ \ \ mass in external legs \ \ \ \\
\hline
&&&\\
\ 1) \ & \  $\alpha(Q^2)$ (ref. \cite{Gossiaux:2008jv}) \
	& $\kappa = 0.2, m_D$ (eq.\ref{prop}) &$ \ \  m_{q,g}=0$\\[1mm]
2) &  $\alpha(Q^2)$ (ref. \cite{Gossiaux:2008jv})  & $\kappa = 0.2, m_D $ (eq.\ref{prop})
	& \ \ $m_{q,g}=m_{q,g}^{DQPM}$ (eq. \ref{equ:Sec2.6})\\[1mm]
3) &  $\alpha(T)$ (eq. \ref{equ:Sec2.9}) & $\kappa = 0.2, m_D$  
	(eq.\ref{prop}) & \ $ m_{q,g}=0$ \\
4) &\  $\alpha(T)$  (eq. \ref{equ:Sec2.9})& $m_g^{DQPM}$ (eq. \ref{equ:Sec2.6}) 
	& \ \ $ m_{q,g}=m_{q,g}^{DQPM}$ (eq. \ref{equ:Sec2.6})\\[1mm]
5) & $\alpha(T)$  (eq. \ref{equ:Sec2.9})& $m_g^{DQPM}$ (eq. \ref{equ:Sec2.6})
	& \ \ $ m_{q,g}=0$\\[1mm]
6) & $\alpha(Q^2)$  (ref. \cite{Gossiaux:2008jv}) & $m_g^{DQPM}$ (eq. \ref{equ:Sec2.6})
	& \ \ $ m_{q,g}=m_{q,g}^{DQPM}$ (eq. \ref{equ:Sec2.6})\\[2mm]
\hline
\end{tabular}

\caption{Coupling constant, gluon masses in the gluon propagator and the masses of the partons in the external legs of the Feynman diagrams for the different curves shown in Fig.\ref{fig:5}.}
\end{table}
%\vspace*{.5truecm}

From Eq. (\ref{eq:x}) one sees immediately that all quantities depend on the distribution of the partonic scattering partners $n_i(k)$. In a thermal medium $n_i(k)$ is given by the Fermi-Dirac, Bose-Einstein or (if quantum statistics is neglected) the Boltzmann distribution. It is obvious that these quantities depend on the local temperature and velocities of the medium, which in the first two approaches are given by a fluid dynamical description of the QGP. In the third approach the evolution is given
by the solution of the Kadanoff Baym equations.
As a consequence, final observables like $R_{\rm AA}$ and $v_2$ are strongly affected by the details of the medium evolution. While the solution of the fluid dynamical conservation equations requires only the knowledge of thermodynamic quantities, such as the equation of state and transport coefficients, the actual nature of the quasiparticles is important for the scattering cross sections between heavy quarks and light partons.
Even if  the Fokker-Planck/Langevin approach does not allow for qualitative comparisons with the data, the drag force is an effective way to compare the stopping of heavy quarks in different transport approaches by reducing the complex kinematics to a function which depends on the temperature and the momentum of the heavy quark only. 

\section{Comparison of the Drag Force for the Different Approaches}

As we have seen, the DQPM embedded in the time evolution of the Kadanoff Baym equation describes very well the
experimental data,  Fig. \ref{raa0}, whereas the effective mass model,  Fig. \ref{fig:K1}, fails. This is astonishing because
the masses of the quarks and gluons are rather similar (Fig. \ref{fig:thmass} and Fig. \ref{fig:thmassdqpm}) and both approaches use Born type diagrams for the interaction of the heavy quarks with the light QGP constituents. 
On the other hand the MC@sHQ, using massless QGP constituents reproduces the data. To elucidate this problem we calculate the drag force for all of the three approaches. In Fig.  \ref{fig:AdragnoK} we compare the drag force of
MC@sHQ approach with the effective mass model, in Fig. \ref{fig:cDragDifferentModel} with the DQPM. In Fig.  \ref{fig:AdragnoK} we see that for larger momenta the energy loss due to radiation becomes more and more dominant
over that due to elastic collisions (coll). We see as well that finite masses (m(T)) reduce the drag force  by large factor, independent of the momentum of the heavy quark, independent of the kind of collisions and also independent of the temperature of the plasma. For a heavy quark with a momentum of 10 GeV this reduction factor is of the order of 4. This is due to the reduced collision rate caused by a lower density of light quarks and gluons at a given temperature and explains why the  MC@sHQ approach with temperature dependent masses fails to describe the data. 
\vspace*{.5truecm}

The comparison between   MC@sHQ  and DQPM is presented in Fig.  \ref{fig:cDragDifferentModel}. The
drag force due to elastic collisions in MC@sHQ is marked by the thin line there. The results of the DQPM is
the  full red line. Regarding the left hand side we observe that at $T=2T_c$ the drag force for DQPM is always smaller then that due to elastic collisions in  MC@sHQ, independent of the heavy quark momentum. So how it is possible that, applied to an expanding plasma, the total energy loss in both approaches is such that the data are reproduced. A first element of the answer is given by the right hand side of Fig. 
 \ref{fig:cDragDifferentModel} which shows the temperature dependence  of the drag force for a heavy quark with a momentum of $p=10\  GeV$. We see that in the DQPM approach the drag force is rather constant below $T=2T_c$ whereas
in MC@sHQ it decreases strongly when approaching the $T_c$.  
What is the origin of this quite different behaviour of  both approaches? This can be inferred from Fig.\ref{fig:5} whose different curves are explained in Table 1. The curve 1) is the drag force for MC@sHQ, curve  4) represents the standard
DQPM calculation.  If we assume for the external legs the quarks masses of DQPM but keep the other parameters like in the original MC@sHQ we obtain curve 2). The difference between 1) and 2) allows for the study of the dependence of the drag force on the parton masses. Curve 2) does not differ substantially from the results presented in Fig. \ref{fig:AdragnoK}. This allows to conclude that finite parton masses lead to a reduction of the drag force and that this reduction increases the closer we come to $T_c$. The form of the drag forces changes completely if replace in  MC@sHQ $\alpha(Q^2)$ by the temperature dependence DQPM coupling constant $\alpha(T)$ (curve 3). In the relevant temperature regime for heavy ion reactions the drag force increases now with decreasing $T/T_c$, means the closer we come to $T_c$ the larger gets the energy loss. If we replace in addition in the gluon propagator the MC@sHQ mass  ($\kappa m_D$) by the PHSD gluon mass (curve 5)  we see
that over the whole temperature range the drag force gets reduced by an almost constant factor as compared to curve 3). 
If one finally takes the DQPM model and replaces only the temperature dependent coupling constant $\alpha(T)$ by that of  
MC@sHQ, which depends on the momentum transfer in the collision $\alpha(Q^2)$, curve 6),  and compares this drag force with that of the standard version of the DQPM, curve 4), one sees the enormous influence on the choice of the coupling constant for the drag force and hence to the energy loss of the heavy quarks close to $T_c$. 
One can conclude from this study that close to $T_c$ finite parton masses at the external legs reduce the drag force whereas
it is increased when employing a temperature dependent coupling constant $\alpha(T)$ instead of $\alpha(Q^2)$. The combination of the DQPM coupling constant and the DQPM masses yields to a less steep decrease of the coupling constant
when the temperature approaches $T_c$ as compared to MC@sHQ.

 Thus MC@sHQ and PHSD (DQPM)  display a quite different scenario for the 
the momentum loss of heavy quarks an a thermal system. In the  MC@sHQ approach the energy loss is much stronger
when the plasma is hot as compared to that close to $T_c$. This means that the energy loss takes dominantly place at the beginning of the expansion whereas in the DQPM approach it is opposite. There close to $T_c$ the energy is almost as large as at high T. It has, however, to be mention that the drag  force in DQPM is for all temperatures lower than the drag for
in  MC@sHQ for elastic collisions only. This means that the nonequilibrium effect which are present in PHSD but not in
MC@sHQ have a very strong influence on the energy loss. 

\section{Summary}
Heavy quarks have been identified as a tool to study the effective degrees of freedom of the QGP. We have studied here three different approaches  - pQCD based MC@sHQ, Dynamical QuasiParticle Model (DQPM) and effective mass approach - where in the last two of them the properties of QGP degrees of freedom (quarks and gluons)
are obtained  by fitting lattice QCD data. We have shown that the presently available experimental data on $R_{AA}$ of  D-mesons can be described by the dynamical models 
(hydro type or transport approach) based on these different propositions of the effective degrees of freedom. That is in spite of strong sensitivity of the drag
force to the model assumptions: finite parton masses and a temperature dependent coupling constant for the heavy quark - light parton collisions modify the drag force in opposite direction, so from momentum loss measurement alone  it will be difficult to disentangle both. In addition, models with a coupling constant which depends on the momentum transfer show the strongest momentum  loss at the beginning of the expansion whereas in those with temperature dependence coupling constant the momentum loss is shifted towards $T_c$. 
In addition, non-equilibrium effects have a strong influence and increase
the energy loss of heavy quarks substantially.  
%This can be attributed to the fact that in expanding system the assumptions on the Einstein relation between longitudinal and %transfers drag coefficient, valid in equilibrium, is no longer validated
%for non-equilibrium situation which is, an addition to the hydrodynamical evolution
%based on local equilibration, end up with much stronger drag force.
It will be subject to a future study to explore this in detail.  

\vspace*{2mm} 
\begin{acknowledgement}
This work was supported by BMBF, by the LOEWE center “HIC for FAIR" and by the project "Together" of the region Pays de la Loire, France.
\end{acknowledgement}

\end{document}